\documentclass[12pt]{article}
\usepackage{amsmath, amsfonts, amssymb, epsfig, palatino, fancyhdr}

\begin{document}

\begin{center}

{\Large \bf Use of G\"{o}del Universe to Construct A New Zollfrei Metric with $R^2 \times S^1$ Topology}

\vspace{1cm}

{\bf Moninder Singh Modgil \footnote[1]{ PhD in Physics, from Indian Institute of Technology, Kanpur, India, and
B.Tech.(Hons.) in Aeronautical Engineering, from Indian Institute of Technology, Kharagpur, India. \\email: moni\_g4@yahoo.com}  }

\vspace{3mm}

{\bf Abstract}

\end{center}

A new example of $(2+1)$-dimensional Zollfrei metric, with the topology $R^2 \times S^1 $, is presented. This metric is readily obtained from the celebrated $(3+1)$- dimensional rotating G\"{o}del universe $G_{3,1}$. This is because $G_{3,1}$ has the interesting property that, the light rays which are confined to move on the plane perpendicular to the rotation axis, return to their origin after a time period $T =  \frac{2 \pi}{\omega}[\sqrt{2}-1]$ -where $\omega$ is the angular velocity of the universe.   Hence by -  the topological identification of  pairs of points on the time coordinate, seperated by the time interval $T$.  and  droping the flat $x_3$ coordinate - which is directed along the rotation axis; one obtains the $(2+1)$-dimensional Zollfrei metric with the  $R^2 \times S^1$ topology.

\vspace{1cm}
\noindent {\bf KEY  WORDS:}  G\"{o}del Universe, Zollfrei Metric, Closed Null Geodesics (CNCs), Closed Timelike Curves (CTCs), Periodic Time

\newpage

 A Lorentzian manifold is said to possess Zollfrei metric, if all its null geodesics are closed \cite{Guillemin 1989}. Part of attraction for studying Zollfrei metrics in physical context comes from the Segal's cosmological model \cite{Segal 1976} which is based upon the $S^3 \times S^1$ topology of spacetime. Examples of manifolds with Zollfrei metric are - $S^n \times S^1$ and $P^n \times S^1$ - with commensurate radii of spatial and temporal factors. To our knowledge, all previously known examples of Zollfrei metric have compact topology for the spatial factor - i.e., $S^n$ or $P^n$. Here, we construct a new example of Zollfrei metric in $(2+1)$-dimensions, with the  $R^2 \times S^1$ topology. This construction is based upon the G\"{o}del universe \cite{Godel 1949}. We shall refer to this as the G\"{o}del-Zollfrei metric, and denoted it by $G_{2,1}^{S^1}$.
 
 The $(3+1)$-D axisymmetric, homogenous, G\"{o}del universe $G_{3,1}$ \cite{Godel 1949} with the topology $R^3 \times R^1$, has the line element -
 
\begin{equation}
ds^2 = a^2 (dx_0^2 - dx_1^2 - dx_3^2 + \frac{e^{2x_1}}{2}dx_2^2 +
2e^{x_1} dx_0 dx_2) \label{Godel line element}
\end{equation}
which satisfies the Einstein equations for a uniform matter density,

\begin{equation}
\rho = (8 \pi G a^2)^{-1},
\end{equation}

\noindent and a cosmological constant,

\begin{equation}
\Lambda
= -4 \pi G \rho,
\end{equation}

\noindent  where, $G$ is Newton's gravitational constant. The universe rotates about the $x_3$ axis, with the angular velocity,

\begin{equation}
 \omega =
2(\pi G \rho)^{1/2} = (\sqrt2 a)^{-1}
\end{equation}

\noindent Lets denote the curved space defined by the three coordinates $(x_0, x_1, x_2)$ as $G_{2,1}$. Now $G_{3,1}$ can be regarded as the product of $G_{2,1}$ and the flat $x_3$ coordinate (which has the topology $R^1$), i.e.,

\begin{eqnarray}
	G_{3,1} = G_{2,1} \times R^1
\end{eqnarray}

Pfarr \cite{Pfarr 1981} worked out geodesic and non-geodesic trajectories for $G_{2,1}$.  For particles confined to move on the $(x_1x_2)$ plane - i.e., the plane perpendicular to the rotation axis $x_3$, he showed that the geodesically moving particles, move on circles and consequently return to the point of their origin. In particular, the light rays move on the largest circles (null geodesics), and reach out to a maximal distance $r_H$, from the point of their origin,

\begin{eqnarray}
	r_H = \frac{2}{\omega} \ln (\sqrt{2}+1)
\end{eqnarray}

\noindent The circle defined by $r_H$, may be termed as the G\"{o}del horizon.   The circles of radius $r=r_H$ are Closed Null Curves (CNCs), while circles of radius $r > r_H$ are Closed Timelike Curves (CTCs). The time taken by the light rays to return to the point of their origin is \cite{Pfarr 1981} -

\begin{eqnarray}
	T = \frac{2 \pi}{\omega}[\sqrt{2}-1]
\end{eqnarray}

\noindent Accordingly if in $G_{3,1}$-
 
\begin{enumerate}

\item One drops the flat $x_3$ coordinate, which is directed along the rotation axis, and

\item Compactifies the time coordinate $x_0$, to $S^1$, and

\item Chooses the time period of $S^1$ time factor equal to $T = \frac{2 \pi}{\omega}[\sqrt{2}-1]$;

\end{enumerate}

\noindent one obtains this new example of $(2+1)$-dimensional Zollfrei metric with the $R^2 \times S^1$ topology.

G\"{o}del \cite{Godel 1949} gave a list of nine interesting properties of $G_{3,1}$. Remarkably, in the fifth property, he considered the possibility of both an open $R^1$ and a closed  $S^1$ time coordinate.

\end{document}